\documentclass[aps,prb,twocolumn,superscriptaddress,longbibliography]{revtex4-2}
\usepackage{mathtools,amssymb,graphicx,units}
\usepackage[usenames,dvipsnames]{color}
\usepackage[plainpages=false,pdfpagelabels,colorlinks=true,linkcolor=blue,urlcolor=magenta,citecolor=magenta,pdftitle={Title},pdfauthor={},pdfdisplaydoctitle=true,pdfduplex=DuplexFlipLongEdge]{hyperref}
\usepackage{subfigure}
\usepackage{grffile}
\usepackage{gensymb}
\usepackage{bm}
\usepackage[version=4]{mhchem}
\usepackage{color}
\usepackage{hyperref}
\usepackage{bibentry}
\usepackage{soul} 

\newcommand{\trt}{TaRhTe$_4$}

\newcommand{\tit}{TaIrTe$_4$}

\begin{document}
	
	\title{Layer dependent topological phases and  transitions in TaRhTe$_4$: From monolayer and bilayer to bulk}

	\author{Xiao Zhang}
	\email{x.zhang@ifw-dresden.de}
	\affiliation{Leibniz IFW Dresden, Helmholtzstr. 20, 01069 Dresden, Germany}
	\affiliation{Institute of Theoretical Physics and W{\"u}rzburg-Dresden  Cluster of Excellence {\it ct.qmat}, Technische Universit{\"a}t Dresden, 01062 Dresden, Germany}
	
	\author{Ning Mao}
	\affiliation{Max Planck Institute for Chemical Physics of Solids, D-01187 Dresden, Germany}
	
	\author{Oleg Janson}
	\affiliation{Leibniz IFW Dresden, Helmholtzstr. 20, 01069 Dresden, Germany}
	
	\author{Jeroen van den Brink}
	\affiliation{Leibniz IFW Dresden, Helmholtzstr. 20, 01069 Dresden, Germany}
	\affiliation{Institute of Theoretical Physics and W{\"u}rzburg-Dresden  Cluster of Excellence {\it ct.qmat}, Technische Universit{\"a}t Dresden, 01062 Dresden, Germany}
	\affiliation{Dresden Center for Computational Materials Science (DCMS), TU Dresden, 01062 Dresden, Germany}
	
	\author{Rajyavardhan Ray}
	\email{r.ray@bitmesra.ac.in}
	\affiliation{Leibniz IFW Dresden, Helmholtzstr. 20, 01069 Dresden, Germany}
	\affiliation{Dresden Center for Computational Materials Science (DCMS), TU Dresden, 01062 Dresden, Germany}
	\affiliation{Department of Physics, Birla Institute of Technology Mesra, Ranchi, Jharkhand, India - 835215}
	\begin{abstract}
		The recently synthesized ternary quasi-2D material TaRhTe$_4$ is a bulk Weyl semimetal with an intrinsically layered structure, which poses the question how the topology of its electronic structure depends on layers separations. 
		Experimentally these separations may be changed for instance by intercalation of the bulk, or by exfoliation to reach monolayer or few-layer structures.
		Here we show that in the monolayer limit a quantum spin Hall insulator (QSHI)  state emerges, employing density functional calculations as well as a minimal four-orbital tight-binding model that we develop.
		Even for weak spin-orbit couplings the QSHI is present, which has an interesting edge state that features Rashba-split bands with quadratic band minima.
		Further we find that a weak topological insulator (WTI) manifests in the bilayer system due to sizable intralayer hopping, contrary to the common lore that only weak interlayer interactions between stacked QSHIs lead to WTIs. 
		Stacked bilayers give rise to a phase diagram as function of the interlayer separation that comprises a Weyl semimetal, WTI and normal insulator phases.
		These insights on the evolution of topology with dimension can be transferred to 
		the family of layered ternary transition metal tellurides.
	\end{abstract}
	
	\maketitle
	
	\section{\label{sec:intro}Introduction}
	
	Topological insulators (TIs) and topological semimetals are new classes of quantum materials \cite{RevModPhys.89.040502,RevModPhys.82.3045,JENature,PhysRevB.83.205101,PhysRevX.5.011029,RevModPhys.83.1057} with potential applications in spintronics and
	quantum computation \cite{Tsai2013,Juan2017,He2022}. TIs
	feature a nontrivial bulk insulating gap accompanied by topologically robust gapless edges for
	two-dimensional (2D) TIs or surface states in three-dimensional (3D) TIs with linearly dispersing
	double degenerate bands meeting at Dirac points \cite{PhysRevB.75.121403,PhysRevB.76.045302,PhysRevB.78.195125,PhysRevB.74.195312}. 2D TIs are characterized by a $\mathcal{Z}_2$ topological index and are also known as quantum spin Hall insulators
	(QSHI) \cite{PhysRevLett.95.226801,PhysRevLett.95.146802,PhysRevB.79.195321,Xiaofeng2014}. For 3DTIs, however, four indices $\mathcal{Z}_2 =(\nu_0; \nu_1\nu_2\nu_3)$ are required based on
	which they can be classified as strong ($\nu_0 = 1$) or weak ($\nu_0 = 0$; $\nu_i =1$ for any
	$i$) \cite{PhysRevLett.98.106803,PhysRevB.75.121306}.
	Correspondingly, strong (weak) TIs possess surface states with an odd (even)
	number of Dirac bands on all (selected) facets \cite{Hsieh2008,PhysRevB.86.045102,Fan2017}. Arguably, the simplest weak TI (WTI) is obtained by stacking of
	QSHIs whereby the facets perpendicular to the stacking direction are free of Dirac nodes (trivial) \cite{PhysRevB.76.045302,LIU2012906}.
	
	Weyl semimetals (WSMs), on the other hand,
	host pairs of topologically protected linear band crossings, called Weyl points (WPs) \cite{PhysRevB.83.205101,PhysRevX.5.011029,Soluyanov2015}. Each WP is
	characterized by a topological charge, chirality $\chi = \pm 1$ \cite{PhysRevLett.107.127205,RevModPhys.93.025002,RevModPhys.90.015001}.
	The surface states of WSMs are Fermi arcs that terminate at the projections of
	bulk WPs of opposite chirality \cite{Jia2016,SYXu2015,PhysRevLett.115.217601}.
	
	In recent years, a large number of TIs and
	WSMs have been experimentally realized, for example, monolayer $1T'$-WTe$_2$ as QSHI \cite{PhysRevX.6.041069,Xiaofeng2014}; Bi$_2$Se$_3$ \cite{Zhang2009,Xia2009} and Bi$_2$Te$_3$ \cite{Hsieh2009,YLChen2009} as 3D strong TIs; ${\rm Bi_{14}Rh_{3}I_{9}}$ \cite{Rasche2013} and ${\rm Bi_{12}Rh_{3}Cu_{2}I_{5}}$ \cite{bric2022} as WTIs; TaAs, \cite{PhysRevX.5.031013,Lv2015} 
	WTe$_2$ \cite{Li2017,PhysRevB.94.121113}, MoTe$_2$ \cite{Deng2016,Jiang2017} and TaIrTe$_4$ \cite{PhysRevB.93.201101, PhysRevB.95.241108} as as WSMs. Many such compounds are layered materials with van der Waals (vdW) type interlayer interactions and exhibit different properties as a function of layer thickness \cite{Otrokov2019,Huang2017,Li2020,Mahatara2021,PhysRevB.102.041109}. Of particular relevance are binary and ternary tellurides known to be WSM in the bulk while the monolayers are QSHIs. 
	From the practical angle, QSHIs with large bulk topological gap for room temperature applications and/or 
	WSMs with minimal symmetry-allowed WPs near the Fermi energy are desired. At the same time, lack of a clear 
	understanding of the topological phase transitions as a function of number of layers is detrimental to possible application in development of efficient quantum devices.
	
	\begin{figure*}
		\centering
		\includegraphics[width=2\columnwidth]{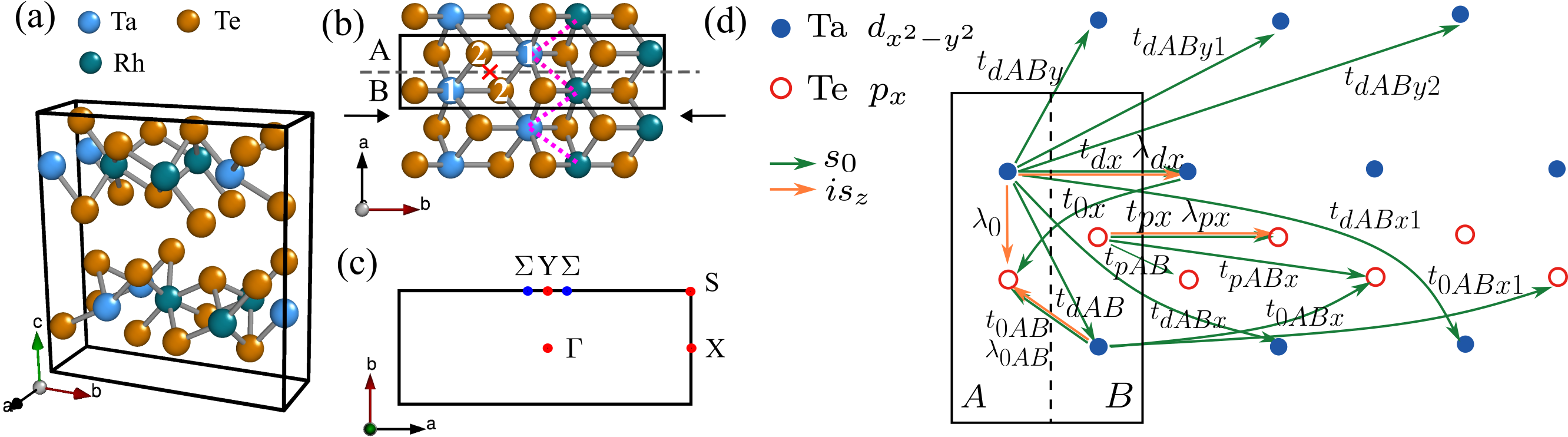}
		\caption{Crystal structures of bulk and monolayer {\trt}.  (a) Crystal structure of bulk and (b) monolayer TaRhTe$_4$ (top view) where the unit cell is shown in solid black lines. In the latter, the sublattices A and B are also shown. The orbitals of Ta(Te)-atoms marked by 1A and 1B (2A and 2B) mainly contribute to minimal tight-binding model. The effective inversion center, marked by a cross, lies at the center of this Ta-Te plaquette. (c) Brillouin zone (BZ) of monolayer TaRhTe$_4$ along with the high symmetry points. 
			(d) Reduced four-site lattice for the tight-binding model with sublattices A and B. The arrows depict the hopping terms ($s_0$) and the (spin-orbit coupling) SOC terms ($s_z$) of the model.} 
		\label{fig1}
	\end{figure*}
	
	Here, we focus on the recently synthesized layered ternary quasi-2D material
	TaRhTe$_4$~\cite{SUFYAN202195,Shipunov2021,D2QI01608G} and elucidate the topological electronic properties of different layered structures, from monolayer and bilayer to bulk
	using density functional theory (DFT). {\trt}  is isostructural to TaIrTe$_4$~\cite{PhysRevB.95.241108,PhysRevB.93.201101,Kumar2021} and
	both share the same nonsymmorphic space group as WTe$_2$~\cite{Soluyanov2015,Li2017,PhysRevB.94.241119}, all of which are type-II
	WSMs in bulk. In comparison, the monolayers of TaIrTe$_4$ and $1T^{\ensuremath{'}}$-WTe$_2$ are QSHIs with time-reversal protected
	edge states~\cite{Yanmeng2019,PhysRevMaterials.3.054206,PhysRevB.99.121105,PhysRevX.6.041069}. Interestingly, the bilayer TaIrTe$_4$, owing to its relatively small interlayer separation, has been predicted to be a QSHI as well~\cite{PhysRevB.102.041109}. With increasing interlayer separation, there is a phase
	transition to a trivial normal insulator (NI) phase. Although remarkable, the QSHI ground state in the bilayer TaIrTe$_4$
	is at odds with the common understanding that interlayer coupling between stacked QSHIs
	gaps out the edge states, leading to a trivial NI phase~\cite{PhysRevLett.107.127205}.   
	Furthermore, it is unknown which topological phases transitions (TPTs) occur while increasing the layer thickness from the monolayers and  bilayers to
	the bulk in any layered chalcogenide, which can be important to bring to fruition the envisioned technological
	applications of these materials.
	
	In this work, we establish that the ambient strucutres of monolayer as well as bilayer {\trt} are QSHI. The 
	monolayer has a bulk gap of $\sim 65$ meV, indicative of a largely insulating behavior at room temperature, while the bilayer 
	is a zero gap semimetal with an average direct gap of $\sim 50$ meV around $\Gamma$.
	We obtain a minimal 4-orbital tight binding (TB) model for the monolayer 
	which reveals that the QSHI phase is driven by hybridization
	between Ta-$5d_{x^2-y^2}$ and Te-$5p_{x}$ orbitals. Our analysis of the interlayer and intralayer hoppings for the
	bilayer QSHI elucidates the importance of interlayer hoppings in the stabilization of the QSHI phase, suggesting that the bilayer {\trt} behaves effectively as a single electronic entity despite vdW separated physical layers in the unit cell.
	
	Finally, the bulk TaRhTe$_4$ is confirmed to be a WSM with three quartet WPs close to Fermi energy arising from band crossings between the top of the valence band and the bottom of the conduction band (bands $N, N+1$ where $N$ is the number of valence electrons). 
	Topological phase transitions between WSM, WTI, and (NI) phase are realized by calibrating the intra- and inter-bilayer
	distances, culminating in a rich phase diagram. 
	The topological phase transition of WTI-NI is the same as that in the bilayer TaRhTe$_4$.
	Phase transitions of WSM-WTI and WSM-NI open up a band gap near the Fermi level and are accompanied by
	creation and annihilation of WPs. Interestingly, we observe that the WPs remain confined in the $k_{z}=0$ plane for WSM-NI as well as WSM-WTI transitions.
	
	\section{\label{sec:monolayer}Monolayer system}
	
	\subsection{Crystal structure and symmetries }
	
	The three-dimensional bulk TaRhTe$_4$ crystallizes in noncentrosymmetric $Pmn2_{1}$ (31) space group, 
	shown in Fig.~\ref{fig1}(a), similar to WTe$_2$ and TaIrTe$_4$. Each unit cell consists of two 
	AB-stacked {\trt} formula units and possess mirror $M_{x}$, glide mirror $\widetilde{M}_{y}=t[(\mathbf{e}_{x}+ \mathbf{e}_{z})/2]M_{y}$, 
	and two fold screw $\widetilde{C}_{2z}=t(\mathbf{e}_{z}/2)C_{2z}$ symmetries. 
	Here, $t(\mathbf{e}_{\mu})$ represents tranlations by a lattice constant along $\mathbf{e}_{\mu}$, $M_{x}$ represents a mirror plane with normal along $\mathbf{e}_x$ direction while $C_{2z}$ represents 2-fold rotational symmetry about the $\mathbf{e}_z$ axis. 
	
	The 2D monolayer {\trt} was constructed out of the bulk crystal structure by inserting large vacuum 
	($\gtrsim 20$ {\AA}) between the AB-stacked layers of the bulk unit cell (see Methods for details).
	Figure~\ref{fig1}(b) shows the top view of monolayer TaRhTe$_4$, Ta and Rh atoms form zigzag chains along 
	the $a$ axis. The resulting structure is dynamically stable [see Supplemental Information (SI) \cite{esi} for the phonon dispersion].
	In comparison to bulk, an effective inversion symmetry can be found in the monolayer TaRhTe$_4$.
	Apart from the time reversal symmetry $\mathcal{T}$ and lattice 
	translations $\{t(\mathbf{e}_{x}),t(\mathbf{e}_{y})\}$, the monolayer structure also has a glide reflection 
	$\widetilde{M}_{x}=t(\mathbf{e}_{y}/2)M_{x}$ and a two-fold screw rotation $\widetilde{C}_{2x}=t(\mathbf{e}_{x}/2)C_{2x}$. 
	The product of these two generators is exactly the spatial inversion symmetry $\widetilde{M}_{x}\widetilde{C}_{2x}=\mathcal{I}$~\cite{PhysRevB.99.121105,PhysRevX.6.041069,PhysRevB.104.035156,PhysRevB.95.241108}. The corresponding inversion center lies at the center of the line joining the Ta atoms. Consequently, the monolayer unit cell can be thought of as composed of sublattices A and B atoms in the unit cell related to each other via the effective inversion symmetry, as shown in Fig.~\ref{fig1}(b). The corresponding BZ along with the high symmetry points are shown in Fig.~\ref{fig1}(c).
	
	\subsection{Electronic properties }
	
\begin{figure}
		\centering
		\includegraphics[width=\columnwidth]{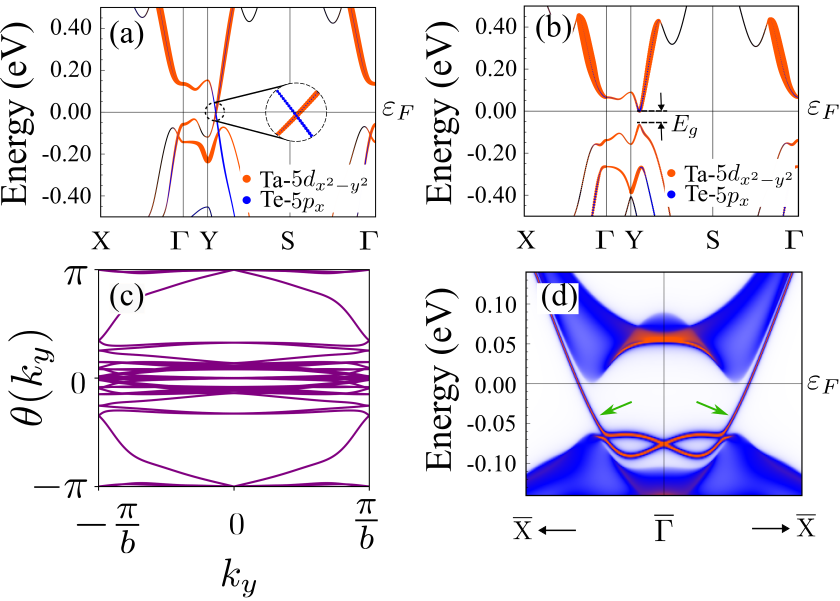}
		\caption{Electronic properties of monolayer {\trt}. Band structure and orbital contributions close to the Fermi energy $\epsilon_{\rm F}$ in monolayer {\trt} (a) without (`no SOC') and (b) with SOC. The inset shows a zoom into the Dirac cone. (c) Wilson loop spectrum for the full DFT band structure when the BZ loop is chosen along $k_x$ direction. (d) Surface spectral function for the termination shown in Fig. \ref{fig1}(b). The edge states are marked by green arrows.}
		\label{fig2}
	\end{figure}

	In Fig.~\ref{fig2}(a), the band structure, obtained within 
	scalar relativistic approximations (`no SOC'), of monolayer TaRhTe$_4$ along a high-symmetry path in BZ is shown. It is a Dirac semimetal with two Dirac nodes located at 
	$\Sigma=(\pm 0.0833,0.253)\,${\AA}$^{-1}$ along Y-S path, {as expected for a non-symmorphic crystal structure}~\cite{PhysRevX.6.041069}. 
	By analyzing the density of states (not shown), we found that Ta-$5d$ and Te-$5p$ orbitals dominate around the Fermi level, whereas the contribution of Rh-$4d$ orbitals is negligible. 
	We thus show the band-weight contributions from Te-$5p_{x}$ and Ta-$5d_{x^2-y^2}$ orbitals in Fig.~\ref{fig2}(a). 
	The overlap of orange and blue dots implies a $p$-$d$ hybridization around the Dirac cone, which is analogous to 
	the case in monolayer TaIrTe$_4$~\cite{PhysRevB.102.041109}. It is important to note that band inversion is present even 
	without SOC, similar to the case of $1T^{\ensuremath{'}}$-WTe$_2$~\cite{PhysRevB.99.121105,PhysRevX.6.041069,PhysRevB.104.035156}.

	As shown in Fig.~\ref{fig2}(b), SOC lifts the degeneracy of the Dirac point and opens up a gap $E_{g} \sim 65$ meV which is 
	sufficiently large for room-temperature QSHI applications. 
	Due to the effective inversion symmetry and time reversal symmetry, each band is doubly degenerate. Further, 
	the Wilson loop spectrum \cite{PhysRevB.84.075119} in Fig.~\ref{fig2}(c) indicates that the bands are topologically 
	nontrivial with the topological invariant $\mathcal{Z}_2 = 1$, implying a QSHI ground state characterized by presence of helical 
	edge states protected by the time-reversal symmetry $\mathcal{T}$.
	
	Figure~\ref{fig2}(d) displays an edge state 
	spectrum of monolayer TaRhTe$_4$ in a nanoribbon geometry for the edge terminations marked by arrows in Fig. 1(b). Indeed, a pair of counter-propagating edge states cross at 
	$\bar{\Gamma}$ and carry opposite spin polarization.
	It is interesting to note that the topological edge Dirac cone lies within the bulk gap and resembles Rashba splitting of edge 
	states with parabolic band minima akin to the nearly free electron dispersion. However, the energy position of the Dirac point is 
	sensitive to the nanoribbon/edge termination 
	[see SI \cite{esi} for details] as well as edge disorder~\cite{PhysRevMaterials.3.054206}.

	\subsection{Four-band minimal tight-binding model }
	
	Given the effective band inversion in monolayer {\trt} even without SOC, the role of SOC in the realization of the QSHI state is {\it a priori} not clear. To further understand the electronic properties of monolayer {\trt}, especially the role of SOC, we obtain a minimal tight-binding (TB) model.
	A minimal TB model is also helpful in investigations of transport properties, studies of disorder effects among others~\cite{Culcer2020,PhysRevMaterials.3.054206}. 
	As shown in the blow-up of Fig. \ref{fig2}(a), the orbital contributions to states close to the Fermi energy are predominantly from Ta-$5d_{x^{2}-y^{2}}$ orbitals and $5p_{x}$ orbitals centered at one of the Te atom marked by atoms 1 and 2 in Fig.~\ref{fig2}(a) and their sublattice symmetry partners ($A/B$). 
	We, therefore, construct a maximally projected Wannier model using a 4-orbital basis set consisting of these orbitals (see Methods for details).
	
	\begin{figure*}
		\centering
		\includegraphics[width=2\columnwidth]{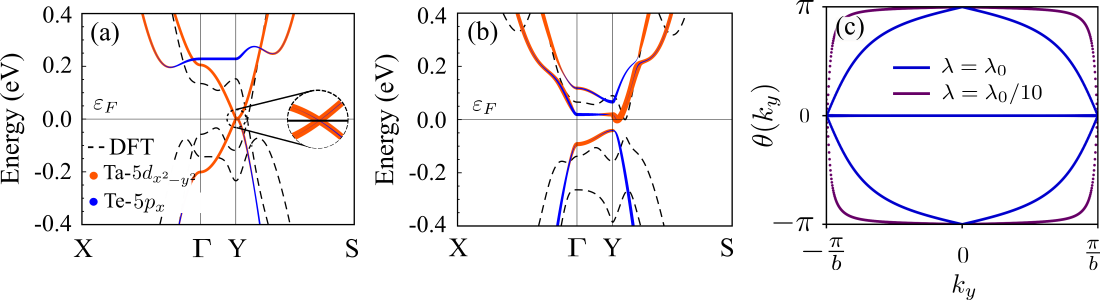}
		\caption{TB model for monolayer {\trt}. Band structures obtained from the 4-orbital TB model for (a) without and (b) with SOC along with the orbital contributions. The DFT bands are shown with black dashed lines for comparison. (c) Wilson loop spectra for the TB model (blue curve), and for the model where the SOC terms are scaled down to 10\% of the real values (purple color).
		}
		\label{fig3}
	\end{figure*}

	Figure~\ref{fig3}(a) shows the orbital-resolved TB band structure without SOC terms. For comparison, the corresponding DFT band structure is also shown (dashed lines). The Wannier TB model successfully reproduces the bulk Dirac cones as well as the band inversion in the vicinity of the Fermi level,
	that are the two characteristic features of the Generalized Gradient Approximation (GGA) band structure within the scalar relativistic approximation (`no SOC'). To reproduce the the Dirac node, however, we need to include hopping terms up to the third-nearest neighbors (up to $\sim 20$\AA). The dominant hopping terms are shown in Fig.~\ref{fig1}(d). 
	We note that, compared to the DFT band structure, the location of the Dirac point in BZ is shifted towards (but not exactly at) the Y point regardless of the number of neighbors considered in the model. The details of the Hamiltonian and the corresponding TB parameters are provided in Methods.

	Upon inclusion of SOC terms which preserve the time reversal symmetry $\mathcal{T}$, glide mirror symmetry $\widetilde{M}_{x}$, glide screw $\widetilde{C}_{2x}$ symmetry, and therefore the effective inversion symmetry $\mathcal{I}$, the resulting TB Hamiltonian matrix is $8\times 8$ and consists of three additional terms (see Methods).
	The corresponding TB band structure along with the DFT band structure (with SOC) is shown in Fig.~\ref{fig3}(b). 
	In the TB and the DFT band structure alike, each band is doubly degenerate. The fundamental gap of minimal model is $\sim 53$ meV, somewhat smaller than the DFT result. However, the band inversion around Y point is reproduced reasonably well. Most importantly, the Wilson loop spectrum in Fig.~\ref{fig3}(c) indicates that the minimal TB model captures the QSHI ground state. 
	
	In order to clarify the role of SOC  in driving the QSHI phase in monolayer {\trt}, we numerically/artificially tune the SOC strength in the 4-orbital minimal TB model. Specifically,  we considered the situation when the effective strength of SOC, $\lambda$, is only 5\% and 10\% of the value in the material-based minimal model, $\lambda_0$ (equivalently, in the minimal TB model with parameters tabulated in Table~\ref{tab:tight_binding_parameters}). Figure~\ref{fig3}(d) shows the Wilson loop calculations for the latter case, $\lambda/\lambda_0 = 0.1$. 
	Remarkably, we find $\mathcal{Z}_{2} = 1$ even in this case, implying that the SOC plays a secondary role in driving the QSHI phase in the monolayer {\trt}.

	\section{\label{sec:bilayer}Bilayer system}
	
	\begin{figure}
		\centering
		\includegraphics[width=\columnwidth]{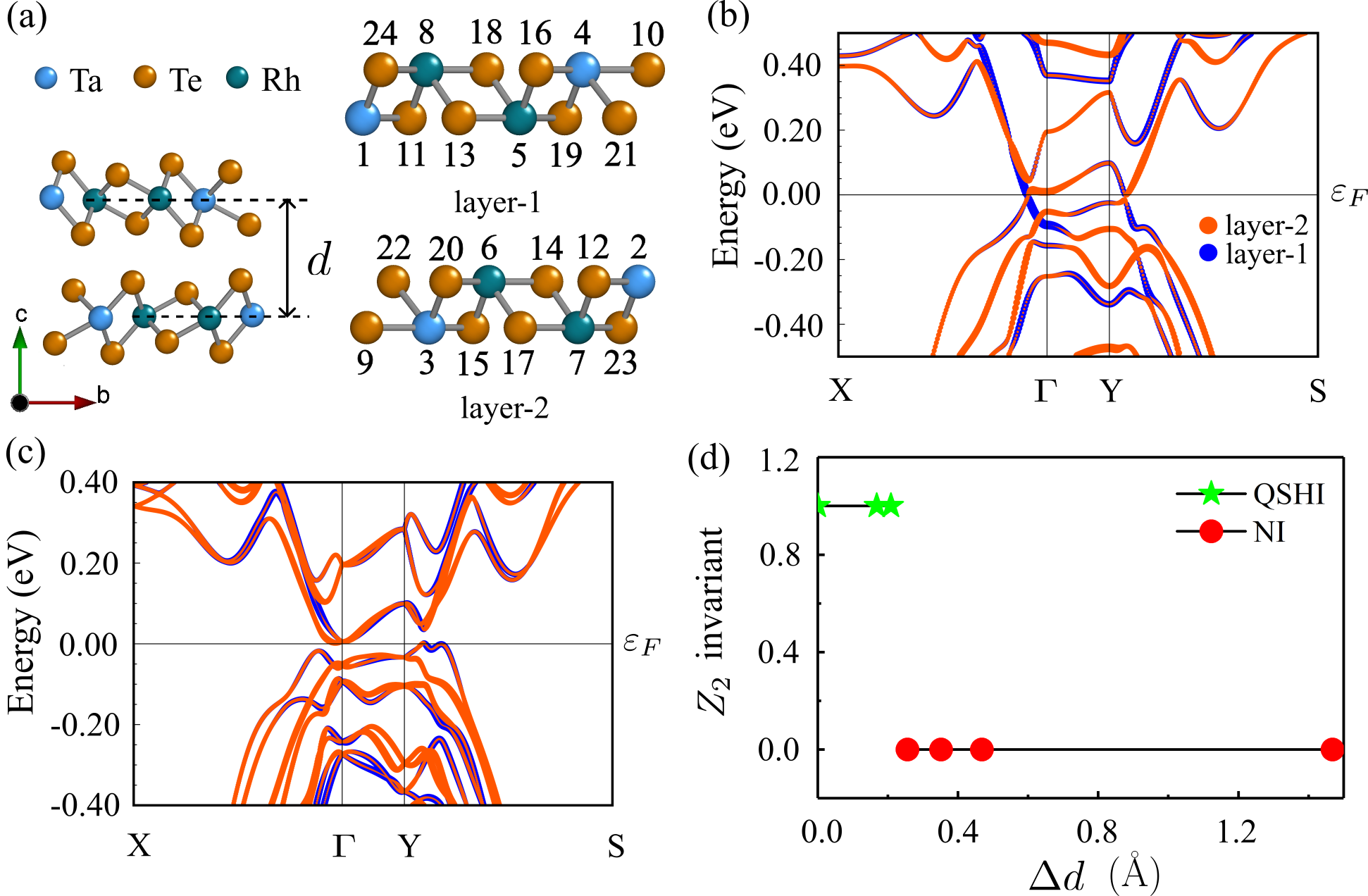}
		\caption{Structure and electronic properties of bilayer {\trt}. (a) Side view of AB stacking TaRhTe$_4$. For ease of discussion, all the atoms have been numbered. The interlayer distance is defined as $d$, where the equilibrium interlayer distance $d_{0}=6.58$ {\AA}. (b) Layer resolved band structure of bilayer TaRhTe$_4$ without SOC. (c) Energy bands of bilayer TaRhTe$_4$ with SOC. (d) Topological $\mathcal{Z}_{2}$ invariant as a function of the interlayer distance $\Delta d$.
		}
		\label{fig4}
	\end{figure}
	
	A bilayer {\trt} system was constructed from the bulk unit cell in a fashion similar to the monolayer (see Methods for details). The unit cell consists of two AB-stacked layers. The structure is shown in Fig. \ref{fig4}(a). Compared to the monolayer, the bilayer breaks the screw symmetry due to the displacement between the layers in the $\mathbf{e}_{x}$-$\mathbf{e}_{y}$ plane. Thus, the effective inversion symmetry is also broken and the space group consists only of mirror reflection $M_x$. 
	
	\subsection{Role of interlayer separation }

	Figure~\ref{fig4}(b) shows the band structure of bilayer TaRhTe$_4$ with layer-resolved orbital contributions along a high symmetry path for the scalar relativistic case (no SOC).
	The highest-lying valence band $N$, where $N$ is the number of valence electrons, and lowest-lying conduction band $N+1$ are quite close along $\Gamma$-X and Y-S but do not cross.
	
	When SOC is included, bands are no longer doubly degenerate, shown in Fig.~\ref{fig4}(c). The ground state is a zero-gap semimetal with direct band gaps around the $\Gamma$ point. The average value of the direct gap is $\sim 50$ meV (min: $\sim 30$ meV, max: $\sim 100$ meV). 
	In contrast to the monolayer case. the band degeneracy is lifted due to the lack of effective inversion symmetry.
	Yet, the bilayer TaRhTe$_4$ also hosts a  nontrivial topological ground state. As the system is quasi-2D but contains two vdW separated layers, the topological index $\mathcal{Z}_2 = (\nu_0; \nu_1 \nu_2 \nu_3)$ based on the Wannier center evolution~\cite{PhysRevB.83.235401} was used (see Methods). 
	$\mathcal{Z}_2 = (0; 0 0 1)$, obtained for the bilayer {\trt}, indicates the presence of edge states on the facets whose normal is perpendicular to the layer direction and corresponds to a QSHI ground state.
	Strictly speaking, the ground state is a WTI, but corresponds to a 2D QSHI due to the quasi-2D structure of the bilayer.
	In SI, we provide an edge state spectrum of bilayer TaRhTe$_4$ in a nanoribbon geometry.

	With increasing interlayer separation $d=d_0 +\Delta d$, where $d_0=6.58$ {\AA} is the interlayer separation in the bulk 3D compound, a topological phase transition to a trivial insulator occurs at $\Delta d = 0.23$ {\AA} ($\Delta d/d_0 = 3.5\%; d=6.81${\AA}); see Fig.~\ref{fig4}(d)). 
	This behavior is similar to the Ir compound, {\tit}, and 
	has been attributed to the strong interlayer coupling~\cite{PhysRevB.102.041109}.

	Interestingly, in {\trt}, the ground state is insensitive to the stacking pattern of the layer in the bilayer unit cell.
	We find that the AA stacked bilayer with no relative displacement in the $\mathbf{e}_{x}$-$\mathbf{e}_{y}$ plane between the layers, and thus retains the screw symmetry and inversion symmetry, also hosts a QSH ground state. 
	The AA stacked TaRhTe$_4$ transitions to a trivial insulator when the interlayer distance is larger than $6.42$ {\AA}. Although the AA stacked structures are deemed unstable~\cite{PhysRevX.6.041069}, efforts are being made to manipulate the interlayer stacking in some transition metal dichalcogenides~\cite{Shinde2018}. This is especially relevant since ternary composition offers a wider possibility to manipulate the composition and structure.
	Additionally, this also allows access to the QSH state in bilayer systems for different interlayer interactions~\cite{PhysRevLett.122.086402}.
	
	\subsection{Intralayer and interlayer hoppings }
\begin{figure}
		\centering
		\includegraphics[width=\columnwidth]{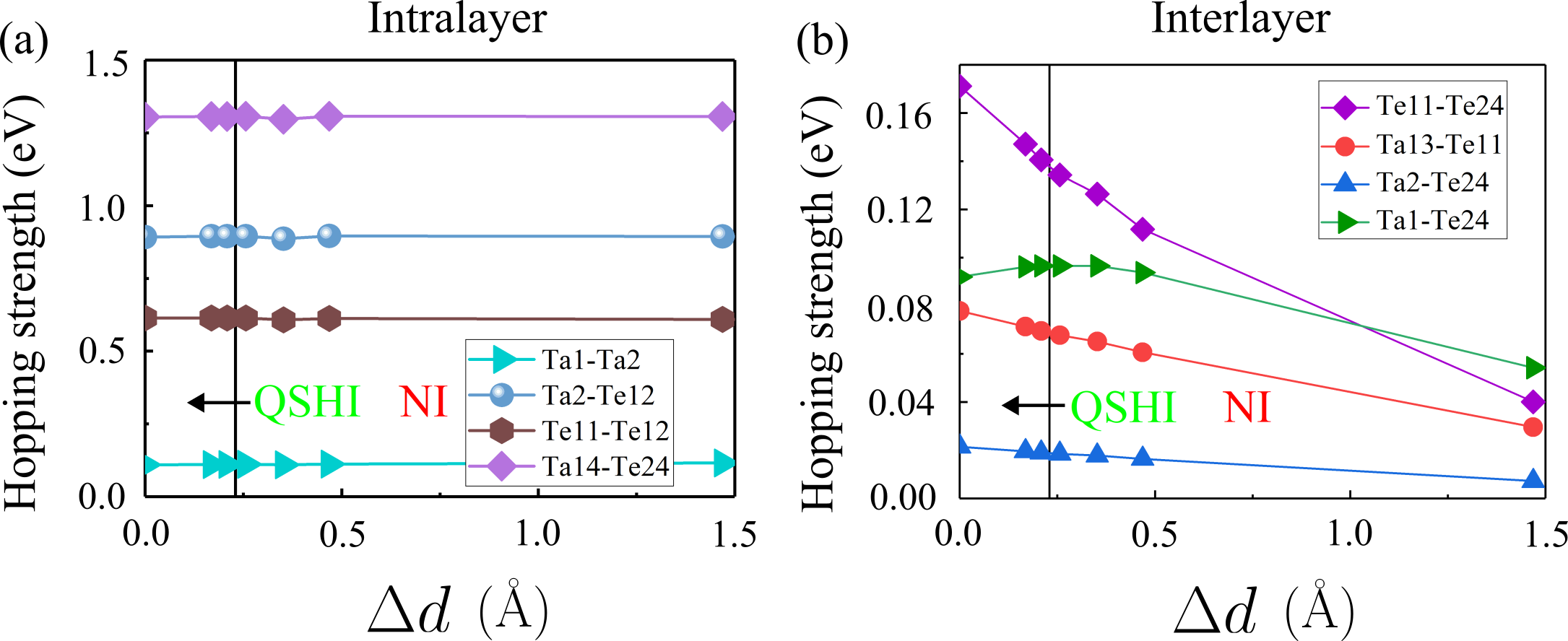}
		\caption{Intra- and interlayer hopping strengths in bilayer {\trt}. Dominant hopping strengths as a function of interlayer distance: (a) intralayer hoppings (b) interlayer hoppings. The atoms in the layer-1 and layer-2 of bilayer structure are marked by numbers in Fig.~\ref{fig4}(a). The vertical black line represents the topological phase transition boundary between QSHI and NI phases.
		}
		\label{fig5}
	\end{figure}

	In order to gain further insights into the topological phase transition in bilayer TaRhTe$_4$, we analyze the dependence of intra- and interlayer hopping parameters on the interlayer distance based on a highly accurate $44$-orbital Wannier model. (see Methods for details). 
	Figure~\ref{fig5}(a) shows selected dominant intralayer hoppings between Ta-Ta and Ta-Te atoms (see SI \cite{esi} for further details). The hoppings between Ta-Te atoms are typically $\sim 1$ eV while Ta-Ta hoppings are an order of magnitude smaller. As could be expected, the intralayer hoppings are almost unchanged with increasing the interlayer separation $d$.
	
	In comparison, the dominant interlayer hoppings range between $\sim 10$ and $\sim 100$ meV , as shown in Fig.~\ref{fig5}(b).
	With increasing interlayer separation $d$, the orbital overlaps decrease and eventually become zero as $\Delta d/d_0 \gg 1$. Some of these terms, in fact, vanish even at $\Delta d \sim 1.5$ {\AA} ($\Delta d/d_0 \sim 18.6\%$). 
	Evidently, the interlayer couplings are reasonably strong to drive the system into a QSHI phase, and decay smoothly across the topological phase boundary.

	The interlayer hopping driven QSHI ground state in bilayer {\trt} is somewhat counter-intuitive as coupling between the Dirac cones from each (mono)layer are expected to gap them out rendering a trivial ground state.
	The large interlayer hoppings are a result of enhanced bandwidths of the monolayer densities of states. 
	The QSHI phase is, however, stabilized by the presence of additional band inversion between the layer-resolved orbital contributions at around Y point, see Fig.~\ref{fig4}(c). As shown in the Supplementary Fig.~S4, 
	with increasing interlayer separation, the contribution by layer-1 to the topmost valence band decreases and the contribution to the lowest valence band increases, while the behavior of layer-2 is just the opposite. Across the phase transition point, the energy gap near the Y point closes and reopens, with significant modifications in the band weight contributions from layer-1 and layer-2. After crossing the phase boundary, the contribution of both the van der Waals separated layers tend to equate in the conduction and valence band edges.
	
	Furthermore, it would be expected that systems with odd number of layers would be QSHIs akin to the monolayer {\trt}. Remarkably, the QSHI bilayer systems implies that few-layer systems with an even number of layers per unit cell would likely also host a QSHI ground state. 
	
	Owing to presence of a robust QSHI ground state in monolayer and bilayer systems, we obtain the exfoliation energies for both of them. 
	It is found that the exfoliation energy for monolayer {\trt} is $2.85$ meV/{\AA}$^2$, while the energy for bilayer system is $1.75$
	meV/{\AA}$^2$ (see SI). For comparison, we also computed the exfoliation energy for graphene in the same way as mentioned in the Methods for reference, and find it to be $1.3$ meV/{\AA}$^2$ comparable to the experimental data~\cite{exfogra}.
	The exfoliation energy of bilayer {\trt} is comparable to that of graphene, while the energy of monolayer system is about 2 times larger than that of graphene, rendering these systems particularly appealing for possible technological applications.

	\section{\label{sec:tpt}Topological phase transitions: from bilayer to bulk }
	
	\subsection{Weyl semimetal phase in 3D TaRhTe$_4$ }
	
	The unit cell of the 3D bulk TaRhTe$_4$ is structurally equivalent to stacked bilayers as shown in Fig.~\ref{fig1}(a). We define the inter-bilayer distance as $c$ and intra-bilayer distance as $d$. The space group and symmetry of the 3D structure have been discussed earlier. 
	The band structure with SOC effects included along high symmetry paths is shown in the Supplementary Fig.~S7. 
	The top of the occupied band (band $N$ where $N$ is the number of valence electrons in the system, blue curve) leads to three-dimensional hole pockets along $\Gamma$-X and $\Gamma$-Y while band $N+1$ (red curve) for electron pocket around $\Gamma$.
	
	The DFT calculations confirm that 3D TaRhTe$_4$ is a Weyl semimetal~\cite{Liu2016}. 
	Within GGA, we obtain three quartets of Weyl points (WPs) between the highest occupied band (blue curve) and the lowest unoccupied band (red curve).
	All the WPs lie in the $k_{z}=0$ plane and are 4-fold degenerate due to symmetry. 
	Table~\ref{tab:wps} lists the position and energy of these WPs alongside their chirality.

	\begin{table}
		\setlength{\tabcolsep}{12pt}
		\centering
		\caption{\small Positions, Chirality ($\chi$), and energies of the Weyl points formed between bands $N$ and $N+1$ in bulk {\trt}, where $N$ is the number of valence electrons. All WPs are four-fold degenerate, related by the symmetries $C_2(z)$ and $M_x$.}
		\begin{tabular}{l r c r}
			\hline\hline
			& $(k_x,k_y,k_z)\frac{2\pi}{a}$ & $\chi$ & $E$ (meV) \\
			\hline
			$W_1$ & $(0.150553, 0.105029,0)$ & $-1$ & $125.2$ \\
			$W_2$ & $(0.019475, 0.098853,0)$ & $+1$ & $-43.0$ \\
			$W_3$ & $(0.025053, 0.066748,0)$ & $+1$ & $-63.7$ \\
			\hline\hline
		\end{tabular}
		\label{tab:wps}
	\end{table}

	\subsection{Phase diagram with intra- and inter-bilayer distance }
	
\begin{figure}
		\centering
		\includegraphics[width=\columnwidth]{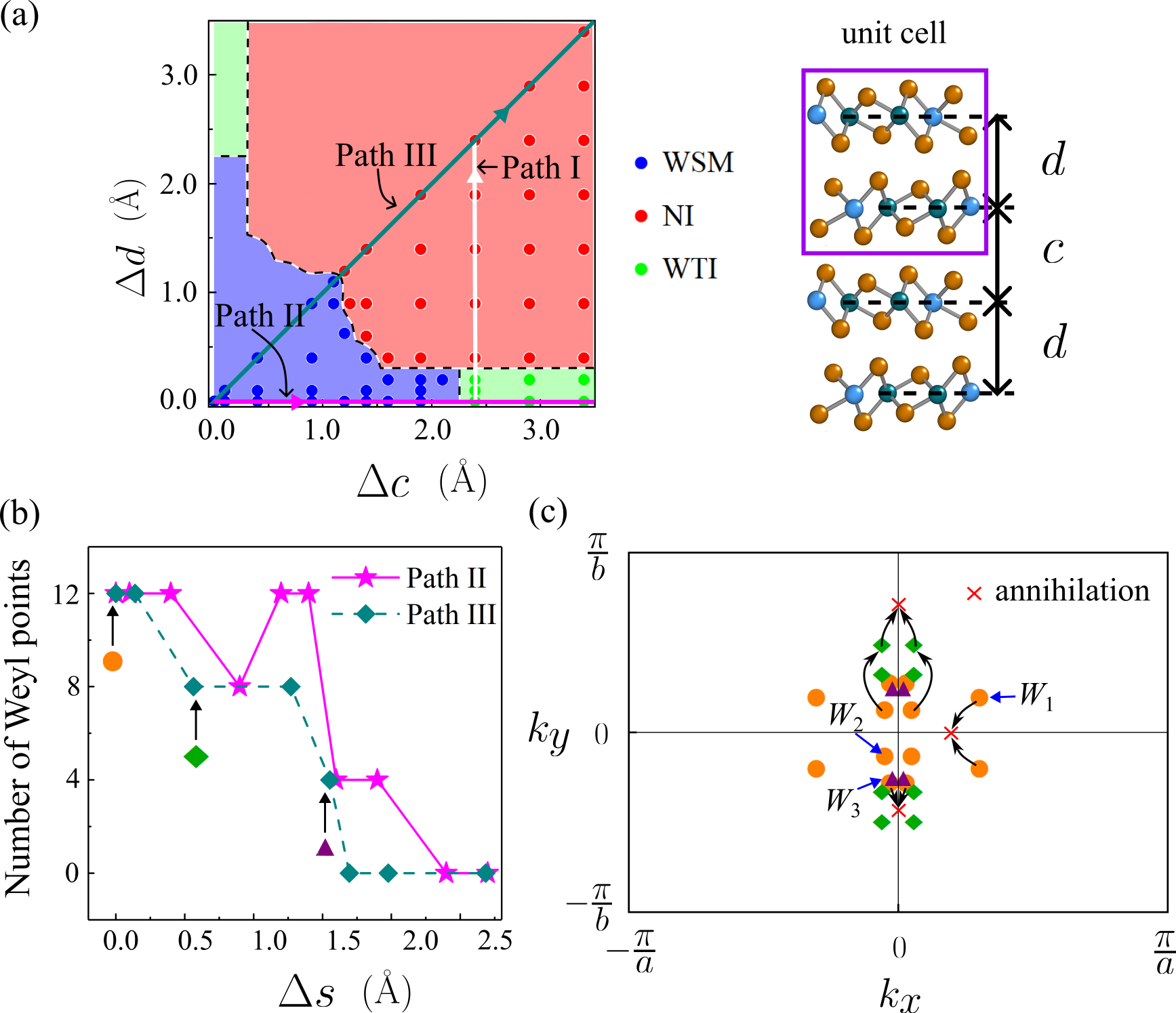}
		\caption{Phase diagram: Topological phase transitions from bilayer to bulk {\trt} as a function inter- and intra-bilayer separations. (a) Left panel: Phase diagram of 3D TaRhTe$_4$ in the $c-d$ plane. Ground state at each point (filled circles) is obtained from DFT calculation. The black dashed curves are a guide to the eye for the phase boundaries. Paths I, II and III highlight the WSM-WTI, WSM-NI, and WTI-NI phase boundaries, respectively. Right panel: side view of \trt with two unit cells, where the inter-bilayer distance $c = c_0 +\Delta c$ and the intra-bilayer distance $d = d_0 +\Delta d$. (b) The number of WPs as a function of $\Delta s$ along Path II and Path III, where $\Delta s=\sqrt{{\Delta c}^2+{\Delta d}^2}$. (c) The coordinates of WPs in BZ selected from Path III of phase diagram. The corresponding chosen points are marked in (b). The red crosses represent the positions in the BZ where the WPs annihilate.
		}
		\label{fig6}
	\end{figure}

	We now turn our attention to possible topological phase transitions from a monolayer and bilayer QSHI to the 3D WSM phase in {\trt}. This is achieved by tuning the intra-bilayer ($d=d_0 + \Delta d$) and inter-bilayer distance ($c=c_0 + \Delta c$), where $c_0= d_0$ represents the interlayer separation for the bulk 3D {\trt} [see the inset of Fig.~\ref{fig6}(a)]. Several interesting possibilities emerge: (i) In the limit of $c,d \rightarrow \infty$, one would expect to recover the topology of the monolayer, while (ii) $c=d=c_0$ represents the bulk 3D system. (iii) $\Delta d=0$, $\Delta c=\infty$ corresponds to a WTI since it can be understood as a stacking of bilayer TaRhTe$_4$ which is a QSHI, along $z$ axis. (iv) $\Delta c \ne \Delta d \ne 0$ represents a generic situation. Due to the large layer separation, the un-/weakly-coupled layers contained in an unit cell are likely to lead to a topological trivial phase. 
	
	We consider a sizable range for the inter- and intralayer separation, up to $\Delta c/c_0 = \Delta d/d_0 = 51.7\%$. It is important to note that while not all the points in the phase diagram may not be experimentally achievable, 
	a sizable region of the phase diagram is relevant for intercalated and/or exfoliated systems for which the TPTs and their boundaries are not known.
	Additionally, the phase diagram must be symmetric about the $c=d$ diagonal and computation of the ground state properties in the bottom triangular region of the $c-d$ phase digram would suffice.

	The resulting phase diagram, shown in Fig.~\ref{fig6}(a), is quite rich and consists of topological insulator (QSHI/WTI), WSM and NI regions. Computations were carried out only in the bottom triangle of phase diagram, because $\Delta c$ and $\Delta d$ are symmetric along the layered direction. All the structures computed are represented by filled circles. The phase boundaries are only a guide to the eye and encompass the phase transitions discussed above.

	To begin with, for $\Delta c/c_0 = \Delta d/d_0 \gg 1$, the unit cell is equivalent to well-separated non-interacting monolayers. The expected ground state is thus a QSHI. From computations of the 3D $\mathcal{Z}_2$ invariant of the considered largest value of $\Delta c/c_0 = \Delta d/d_0 = 51.7\%$, however, we find a trivial NI ground state. 
	On the other hand, for $\Delta d=0$ and $\Delta c/c_0 \gg 1$  corresponds to a bilayer system and the ground state is a WTI as discussed earlier. For such a suitable but fixed $\Delta c/c_0$, varying $\Delta d$ (Path I in Fig.~\ref{fig6}(a)) leads to a topological phase transition between the WTI and the NI phase, {\it i.e.} from $\mathcal{Z}_2 = (0;001)$ to $\mathcal{Z}_2 = (0;000)$ for $\Delta c=2.4$ {\AA} and $\Delta d=0.23$ {\AA}, {\it i.e.}, $ \Delta c/c_0= 36.5\%$ and $\Delta d/d_0= 3.5\%$, and was discussed above in detail.

	
	Similarly, for $\Delta c = \Delta d = 0$, corresponding to the bulk 3D {\trt}, the ground state is a WSM. Keeping $\Delta d =0$ fixed, and increasing $\Delta c/c_0$ corresponds to a structural phase transition from the bulk 3D structure to the AB-stacked bilayer {\trt}. By the same token, a topological phase transition from the WSM phase to the WTI ground state is found, as shown in Path II along $\Delta d = 0$.
	
	The WSM phase extends to regions of small $\Delta c$ and $\Delta d$, whereas WTI is located in the area with large $\Delta c$ but small $\Delta d$. With large $\Delta c/c_0$ and $\Delta d/d_0$, the {\trt} system becomes a NI.
	Naturally, band gap opens when the phase transitions of WSM-WTI and WSM-NI occur.
	At the same time, the inter- and intra-bilayer separation induces creation and annihilation of WPs in the WSM phase.

	To further illustrate these topological phase transitions, we consider paths II and III which, respectively, highlight the WSM-WTI and WSM-NI topological phase transitions.
	Figure~\ref{fig6}(b) shows the number of Weyl points as a function of $\Delta s$ along these paths. 
	Because of the time-reversal symmetry, Weyl nodes are always created and annihilated as quartets. 
	Along Path II, the number of WPs in TaRhTe$_4$ decreases to 4 before the phase transition, and completely annihilated at around $\Delta s = \sqrt{\Delta c^2  + \Delta d^2} \sim 2\,${\AA}. 
	
	Similar behavior can be observed along Path III. 
	A band gap opens up already at around $\Delta s \sim 1.7$ {\AA}. 
	Due to $c=d$, all symmetries in equilibrium 3D structure of bulk {\trt} are preserved. 
	Therefore, all the Weyl points are still constrained in $k_{z}=0$ plane. 
	In Fig.~\ref{fig6}(c), we show the evolution of the WPs along path III. For brevity, we depict the coordinates of WPs for selected values of $\Delta s$. the three quartet of WPs found in the bulk {\trt}, for $\Delta s = 0$, are denoted by orange dots. Their positions in the BZ are listed in Table \ref{tab:wps}. As $\Delta s$ increases, the quartet $W_1$ moves continuously closer to the origin and eventually annihilate each other on the $k_x$ axis. Concurrently, the WPs $W_2$ and $W_3$ move along the $k_y$ axis increasing their relative separations. At $\Delta s = 0.5\,\text{\AA}$, only two quartets ($W_2$ and $W_3$) survive and their positions in the BZ is shown by green diamonds. Further increasing $\Delta s$ creates further WPs [see Fig.~\ref{fig6}(b)] close to the $k_x$ axis which annihilate rather quickly with small changes in $\Delta s$ (not shown). Simultaneously, both quartet $W_2$ and $W_3$ also get annihilated. Increasing $\Delta s$ even further to $\Delta s = 1.7\,\text{\AA}$ creates a new quartet of WPs close to $k_x = 0$ axis, shown by purple triangles which eventually annihilate each other at the phase boundary between the WSM and the NI states.

	\section{\label{sec:outlook}Conclusion \& Outlook}
	
	Our investigation of the monolayer, bilayer and bulk {\trt} systems using DFT brings forth several interesting and important aspects of layered vdW tellurides in general and {\trt} in particular.
	First, monolayer {\trt} is a QSHI with 65 meV bulk gap, suggesting viability for room temperature applications. At the same time, the minimal 4-orbital tight binding model presented here will enable further investigations of disorder and geometric effects for realistic samples.
	
	Second, our analysis of the inter- and intralayer TB parameters clarifies that, even if vdW separated, the bilayer effectively behaves as a single electronic unit and hosts a QSHI ground state. Presence of QSHI ground state for both monolayer and bilayer, as well as the insensitivity of the QSHI ground state to the stacking pattern in the bilayer, suggests that even few layer {\trt} systems would also be QSHIs, and may also have interesting consequences for several related compositions and compounds.
	
	Third, and perhaps the most important, a rich phase diagram containing QSHI/WTI, WSM and NI phases is obtained by varying the interlayer and inter-bilayer separations, comprehensively cover a vast majority of possible realistic situations under different synthesis conditions.

	These insights are easily extendable not only to multi-/few-layer {\trt} but to binary and ternary tellurides in general, and intercalated and exfoliated tellurides in particular.  Moreover, the higher degree of control offered by ternary compounds in terms of their composition and structure compared to their binary counterparts suggest that these our findings will likely impact the field of 2D chalcogenides significantly and drive not only future theoretical and experimental investigations, but also open up new avenues for quantum devices.

	\section{Methods}
	
	\subsection{Crystal structures } The bulk crystal structure consists of four formula units and two vdW separated layers per unit cell. The lattice constants are $a=3.76$ {\AA}, $b=12.55$ {\AA}, and $c=13.17$ {\AA}~\cite{Shipunov2021}. The interlayer separation is $c_0 = 6.58\,\text{\AA}$. The optimized atomic positions were used while the lattice constants were kept fixed at experimental values~\cite{Shipunov2021}.

	Monolayer {\trt} was constructed from the bulk unit cell by inserting a large vacuum between the two layers in the bulk unit cell. It was found that a vacuum  $\gtrsim 15\, \text{\AA}$ does not influence the electronic properties of the monolayer. The atomic positions of monolayer structure were first optimized under $Pmn2_{1}$ symmetry such that the force on each atoms is less than 1 meV/{\AA}. Then the lattice constants were tuned to minimize the total energy of the system. At last, the atomic positions were optimized again under the new lattice constants.
	The in-plane lattice parameters are $a=3.72$ {\AA} and $b=12.42$ {\AA}, 
	which are slightly smaller than bulk values.
	The electronic bands of final structure matches the results by using the structure from Materials Cloud~\cite{matcld}. For comparison, the energy bands with optimized atomic positions but unoptimized lattice constants are presented in SI~\cite{esi}.

	Similarly, bilayer {\trt} systems were prepared by inserting large vacuum between the bilayers in the bulk unit cell. For the bilayer system with intra-bilayer distance equal to the interlayer separation in the bulk 3D system, but large inter-bilayer distance, the atomic positions were optimized with a force threshold of 1 meV/{\AA} for each atom.
	Other bilayer structures with different intra-bilayer separations realized by inserting the vacuum directly in the optimized structure.
	The resulting crystal structures for monolayer as well as bilayer systems are presented in SI~\cite{esi}. 
	
	\subsection{DFT calculations }
	The DFT calculations were carried out for monolayer, bilayer and bulk {\trt} systems using the Full-Potential Local-Orbital (FPLO) code~\cite{PhysRevB.59.1743}, version 18.57~\cite{fplo}. The Perdew-Burke-Ernzerhof implementation~\cite{pbe} of the GGA was employed in all cases. For the numerical integration of monolayers and bilayers, we use a  $k$ mesh with ($20\times 10 \times 1$) intervals in the BZ, and a $k$ mesh with ($20\times 10 \times 10$) intervals for bulk TaRhTe$_4$. The linear tetrahedron method was used for the numerical
	integration in the Brillouin zone.
	As applicable, the scalar
	relativistic correction (`no SOC') was used, while the SOC effects (`with SOC') were included via the 4-spinor formalism as implemented 
	in the FPLO code. 
	
	\subsection{Topological electronic properties }
	For the monolayer insulators, a highly accurate maximally projected Wannier model was constructed which was used for the computation of the topological properties. It was constructed by projecting the band structure onto a subset of atomic-like basis functions (Wannier functions) within a given energy window. 
	We focused on a set of isolated bands across the Fermi energy, lying between $-6.5$ eV and $+2.3$ eV, which have dominant contributions from the Ta-$5d$ orbitals, Rh-$4d$ orbitals, and Te-$5p$ orbitals.
	
	The topology of the band structure was determined by calculating the $\mathcal{Z}_2$ index with via the Wilson loop method (own code)~\cite{PhysRevB.84.075119}. Sensitivity of the ground state to the Te positions was explicitly checked by considering an inversion symmetric unit cell with relaxed atomic positions for which the $\mathcal{Z}_2$ index matches with the corresponding value obtained via the method of Wannier center evolution provided by FPLO, suggesting the topology is robust against perturbations to Te positions.
	
	The edge state spectra were obtained by computing the bulk projected bands using nanoribbon geometry as implemented in \textsc{pyfplo} module of the FPLO code. Two blocks of unit cell are used as ribbon width in all semi-slab calculations. 200 points were sampled across the energy window spanning $[-0.15, 0.15]$ eV.
	
	For the bilayers, with direct gap throughout the BZ, the topological properties were ascertained from the 
	$\mathcal{Z}_2=(\nu_0; \nu_1 \nu_2 \nu_3)$ index for the 3D systems, obtained via the Wannier center evolution 
	algorithm~\cite{PhysRevB.83.235401} as implemented in the FPLO code. 
	
	The Wannier tight-binding model comprises the Ta-$5d$ orbitals, Rh-$4d$ orbitals, and Te-$5p$ orbitals with 
	energy window lying between $-7.0$ eV and $+2.5$ eV. The basis set for the Wannier projections consists of 88 orbitals 
	for the scalar relativistic case (`no SOC'). Correspondingly, the basis of Wannier functions involved 176 orbitals 
	when spin-orbit interactions were considered. Because varying the interlayer separations does not significantly change 
	the distribution of bands, the energy window was kept between $-7.0$ eV and $+2.5$ eV in all bilayer computations.
	
	For the bulk {\trt}, on the other hand, the ground state is semimetallic. 
	To study its underlying topology of the
	electronic structure, we constructed a Wannier model projecting the band structure onto Wannier functions for the isolated manifold of bands between $-6.5$ eV and $+4.3$ eV, which have dominant
	contributions from the Ta-$5d$, Rh-$4d$, $5s$, and
	Te-$5p$ orbitals. Therefore, the basis set for the Wannier
	projections consists of 92 orbitals for the scalar relativistic case. Correspondingly, the
	basis of Wannier functions involved 184 orbitals 
	when spin-orbit interactions were considered.
	The WPs were obtained and confirmed by computing the Chern
	numbers as implemented in \textsc{pyfplo} module of the FPLO code, and outlined
	in Ref.~\cite{PhysRevB.93.201101}.
	
	\subsection{Exfoliation energy }
	In order to estimate the energy cost of exfoliating monolayer and bilayer {\trt} from a real bulk system, we calculated their exfoliation energies. We define exfoliation energy as the energy needed to completely detach all the layers in the bulk unit cell. It is worth noting that in certain studies~\cite{Wang2015,PhysRevB.85.205418}, exfoliation energy is alternatively defined as the energy expended in removing the top layer from the surface of a bulk crystal. Thereby, we compute the total energy of the bulk {\trt} as a function of the intra-bilayer distance $d$ and inter-bilayer distance $c$, and obtain the difference from the energy of the equilibrium structure. For bilayer system, we vary the interlayer distance $c$ only. To arrive at monolayer limit, we vary both the intra-bilayer and inter-bilayer distances at the same time. For these calculations, we use a $k$ mesh with ($20\times 20\times 10$) intervals in the BZ. 
	Because there are 2 layers per unit cell, we divide the energy by 4 times surface area ($3.76\times 12.55\;\mathrm{\AA}^2$) of unit cell as the final monolayer exfoliation energy, whereas divide 2 times surface area as the final bilayer exfoliation energy. As a benchmark, we also obtain the exfoliation energy for graphite. The situation of graphene is same as that of the monolayer, {\it i.e.} we divide the energy by by 4 times surface area to obtain the exfoliation energy.
	
	\begin{table*}
		\setlength{\tabcolsep}{12pt}
		\centering
		\caption{\small Tight-binding model parameters of the 4-orbital minimal model for monolayer {\trt}. The model parameters correspond to Eq. (\ref{eq1}) and Eq. (\ref{eq2}) and are shown in Fig. \ref{fig1}(d).}
		\begin{tabular}{l r l r}
			\hline\hline
			{\bf Parameter} & {\bf Value (meV)} & {\bf Parameter} & {\bf Value (meV)} \\
			\hline\hline
			\multicolumn{4}{c}{\textbf{A. TB model without SOC $H_{0}$}} \\
			\hline 
			$\mu_p$ & $590\,$ & $t_{dABx1}$ & $30\,$ \\
			$\mu_d$ & $-2070\,$ & $t_{dABy}$ & $-26\,$ \\
			$t_{px}$ & $940\,$ & $t_{dABy1}$ & $-11\,$ \\
			$t_{dx}$ & $-290\,$ & $t_{dABy2}$ & $-15\,$ \\
			$t_{pAB}$ & $240\,$ & $t_{0AB}$ & $-120\,$  \\
			$t_{pABx}$ & $-38\,$ & $t_{0ABx}$ & $-81\,$  \\
			$t_{dAB}$ & $-110\,$  &  $t_{0ABx1}$ & $-26\,$  \\
			$t_{dABx}$ & $27\,$  & $t_{0x}$ & $57\,$     \\
			\hline 
			\multicolumn{4}{c}{\textbf{B. SOC terms $H_{\rm SOC}$}} \\
			\hline 
			$\lambda_{0AB}^z$ & $16.8\,$ & $\lambda_{px}^z$ & $7.5\,$ \\
			$\lambda_{dx}^z$ & $7.1\,$ & $\lambda_{0}^{z}$ & $9.8\,$\\
			\hline\hline
		\end{tabular}
		\label{tab:tight_binding_parameters}
	\end{table*}

	\subsection{Phonon spectrum }
	
	We calculate the phonon spectrum of monolayer {\trt} by using Vienna ab initio Simulation Package (VASP)~\cite{PhysRevB.54.11169}, version 5.4.2, and {\sc phonopy} package~\cite{PhysRevB.78.134106}, version 2.20.0. The second-order harmonic interatomic force constants are calculated by the density-functional perturbation theory with a ($2 \times 2 \times 1$) supercell and ($5\times 2\times 1$) $k$ mesh, employing projector augmented-wave pesudopotential~\cite{PhysRevB.50.17953}, where the cutoff energy is set as $500$ eV. In SI, we show the phonon spectrum of monolayer {\trt}.

	\subsection{Minimal TB model for monolayer {\trt} }
	
	For the monolayer, a 4-orbital TB model was also obtained. Starting from the symmetry based Hamiltonian in Ref.~\onlinecite{PhysRevMaterials.3.054206} and considering long range hopping terms, up to $\sim 20${\AA}, the Hamiltonian without SOC is written as:
	\begin{align}\label{eq1}
		\begin{split}
			H_{0}=&\left[\frac{\mu_{p}}{2}+t_{px}\cos(k_{x}a)\right]\Gamma_{1}^{-}
			+\left[\frac{\mu_{d}}{2}+t_{dx}\cos(k_{x}a)\right]\Gamma_{1}^{+}\\
			&+\left[t_{dAB}(1+e^{ik_{x}a})
			+t_{dABx}(e^{-ik_{x}a}+e^{2ik_{x}a})\right.\\
			&+t_{dABx1}(e^{-2ik_{x}a}+e^{3ik_{x}a})
			+t_{dABy}e^{-ik_{y}b}(1+e^{ik_{x}a})\\
			&+t_{dABy1}e^{-ik_{y}b}(e^{-ik_{x}a}+e^{2ik_{x}a})\\
			&\left.+t_{dABy2}e^{-ik_{y}b}(e^{-2ik_{x}a}+e^{3ik_{x}a})\right]e^{i\mathbf{k}\cdot\Delta_{1}}\Gamma^{+}_{2}\\
			&+\left[t_{pAB}(1+e^{ik_{x}a})
			+t_{pABx}(e^{-ik_{x}a}+e^{2ik_{x}a})\right]e^{i\mathbf{k}\cdot\Delta_{2}}\Gamma_{2}^{-}\\
			&+\left[t_{0AB}(1-e^{ik_{x}a})+t_{0ABx}(e^{-ik_{x}a}-e^{2ik_{x}a})\right.\\
			&\left.+t_{0ABx1}(e^{-2ik_{x}a}-e^{3ik_{x}a})\right]e^{i\mathbf{k}\cdot\Delta_{3}}\Gamma_{3}\\
			&-2it_{0x}\sin(k_{x}a)\left[e^{i\mathbf{k}\cdot\Delta_{4}}\Gamma_{4}^{-}+e^{-i\mathbf{k}\cdot\Delta_{4}}\Gamma_{4}^{+}\right]
			+\text{H.c.}
		\end{split}
	\end{align}
	These hopping terms are illustrated in Fig.~\ref{fig1}(d).
	The $4\times 4$ matrices $\Gamma_{i}$ are linear combinations of products $\tau_{i}\sigma_{i}$ of Pauli matrices and $\Delta_{i}$ are vectors corresponding to interatomic distances and is defined below.
	
	Specifically, the matrices $\Gamma_i$ have the form:
	\begin{eqnarray}
		\Gamma_0 &=& \tau_0\sigma_0,\\
		\Gamma_1^\pm &=& \frac{\tau_0}{2}(\sigma_0 \pm \sigma_3),\\
		\Gamma_2^\pm &=& \frac{1}{4}(\tau_1 + i\tau_2)(\sigma_0 \pm \sigma_3),\\
		\Gamma_3 &=& \frac{1}{2}(\tau_1 + i\tau_2)i\sigma_2, \\
		\Gamma_4^\pm &=& \frac{1}{4}(\tau_0 \pm \tau_3)(\sigma_1 + i\sigma_2),\\
		\Gamma_5^\pm &=& \frac{\tau_3}{2}(\sigma_0 \pm \sigma_3),\\
		\Gamma_6 &=& \frac{1}{2}(\tau_1 + i\tau_2)\sigma_1,
	\end{eqnarray}
	where the matrices $\tau_j\sigma_i$ are products of Pauli matrices acting in orbital space with respect to the basis $\lbrace d_{\mathbf{k}Ads}, d_{\mathbf{k}Aps}, d_{\mathbf{k}Bds}, d_{\mathbf{k}Bps}\rbrace$, where $d_{\mathbf{k}cls}$ annihilates an electron with momentum $\mathbf{k}$, spin-$S_z$ eigenvalue $s=\uparrow,\downarrow$ and orbital $l=p,d$ (Te,Ta) in sublattice $c=A,B$.
	In other words, $\tau_j$ acts on the sublattice degree of freedom and $\sigma_i$ acts on the orbital degree of freedom.
	
	The vectors in unit cell are defined as $\Delta_{1}=\mathbf{r}_{Ad}-\mathbf{r}_{Bd}$, $\Delta_{2}=\mathbf{r}_{Ap}-\mathbf{r}_{Bp}$, $\Delta_{3}=\mathbf{r}_{Ad}-\mathbf{r}_{Bp}$, and $\Delta_{4}=\mathbf{r}_{Ad}-\mathbf{r}_{Ap}$, where the vector $\mathbf{r}_{cl}$ denotes the position of the corresponding lattice site, associated with an orbital $l$ in sublattice $c$, in the unit cell. We list the coordinates of vectors as $\mathbf{r}_{Ad}=(-0.25a, 0.16b)$,  $\mathbf{r}_{Bp}=(0.25a, 0.044b)$, $\mathbf{r}_{Ap}=(-0.25a, -0.044b)$, and $\mathbf{r}_{Bd}=(0.25a, -0.16b)$.
	
	When taking into account SOC, the minimal model is constructed by localized Wannier orbitals of the Ta-$d$ and Te-$p$ orbitals along with the spin degrees of freedom, which gives rise to an eight-orbital model. Based on the minimal tight-binding model without SOC, we include spin-orbit coupling terms that preserve: time reversal symmetry $\mathcal{T}$, glide mirror symmetry $\widetilde{M}_{x}$, glide screw $\widetilde{C}_{2x}$ symmetry, and inversion $\mathcal{I}$ symmetry. In the basis of the Hamiltonian in Eq.~(\ref{eq1}), the matrix representations of the symmetry operations are $\hat{\mathcal{T}}= is_{y}\Gamma_{0}K$ with complex conjugation $K$, $\hat{\mathcal{I}}=s_{0}\tau_{1}\sigma_{3}$, and $\hat{M}_{x} =is_{x}\tau_{0}\sigma_{3}$ for the nontranslational component of $\widetilde{M}_{x}$. By taking the hopping parameters ($\geqslant 0.014$) eV from Wannier fit result, the Bloch Hamiltonian of the final tight-binding model with SOC is $H(\mathbf{k})=s_{0}H_{0}(\mathbf{k})+H_{\text{SOC}}(\mathbf{k})$, with
	\begin{align}\label{eq2}
		\begin{split}
			H_{\text{SOC}}=&\lambda_{dx}^{z}s_{z}\sin(ak_{x})\Gamma_{5}^{+}+\lambda_{px}^{z}s_{z}\sin(ak_{x})\Gamma_{5}^{-}\\
			&-i\lambda_{0AB}^{z}s_{z}(1+e^{iak_{x}})e^{i\mathbf{k}\cdot\Delta_{3}}\Gamma_{6}\\
			&-i\lambda_{0}^{z}s_{z}(e^{i\mathbf{k}\cdot\Delta_{4}}\Gamma_{4}^{+}-e^{-i\mathbf{k}\cdot\Delta_{4}}\Gamma_{4}^{-})+\text{H.c.}
		\end{split}
	\end{align}
	where $s_i$ are Pauli matrices acting in spin space and $s_{0}$ is the $2\times 2$ identity matrix. The full Hamiltonian is given by $H = H_0 + H_{\rm SOC}$. Parameter values in Eqs.~(\ref{eq1}) and (\ref{eq2}) are listed in Table~\ref{tab:tight_binding_parameters}.


	\acknowledgements
	
	The authors thank Hui Liu and Ion Cosma Fulga for helpful discussions and Ulrike Nitzche for technical assistance. RR acknowledges helpful discussions with 
	Anupam Roy. 
	We acknowledge financial support from German Forschungsgemeinschaft (DFG, German Research Foundation) via SFB1143 Project No. A05 and 
	under Germany's Excellence Strategy through Würzburg-Dresden Cluster of Excellence on Complexity and Topology in Quantum Matter - ct.qmat 
	(EXC 2147, Project No. 390858490).
	
	
	%

	\clearpage
	
	\onecolumngrid
	
	\pagebreak
	
	\clearpage
	
	\pagenumbering{arabic}
	\section*{Supplemental Information}
	\renewcommand{\thesection}{\Alph{section}}
	\setcounter{section}{0}
	\renewcommand{\theequation}{S\arabic{equation}}
	\setcounter{equation}{0}
	\renewcommand{\thefigure}{S\arabic{figure}}
	\setcounter{figure}{0} 
	\renewcommand{\thetable}{S\arabic{table}}
	\setcounter{table}{0} 
	
	In Supplementay~\ref{sec:App_str}, we provide the details of the relaxed crystal structures for monolayer, bilayer and bulk {\trt} used in this study. Supplementay~\ref{sec:App_mband} illustrates the DFT band structure of monolayer {\trt} which lattice constants are same to the bulk. Supplementay~\ref{sec:exfoliation} shows the exfoliation energy of monolayer and bilayer {\trt}. In Supplementay~\ref{sec:App_phono}, the phonon spectrum of monolayer {\trt} is provided. Supplementay~\ref{sec:bilayer_inversion} gives the layer-resolved band weight of bilayer {\trt}. Supplementay~\ref{sec:App1} presents the surface spectral function for different edge terminations in the monolayer {\trt}. Supplementay~\ref{sec:AAstack_bands} shows bandstructure and Wilson loop of AA-stacked bilayer {\trt}, while Supplementay~\ref{sec:App_bulk_bands} shows the DFT bandstructure for bulk {\trt}.

	\section{\label{sec:App_str}Crystal structures of monolayer, bilayer, and bulk {\trt}.}

	\begin{table}[h!]
		\setlength{\tabcolsep}{20pt}
		\begin{center}
			\caption{Crystal structure of monolayer TaRhTe$_4$.  It was constructed from the bulk unit cell by inserting a large vacuum between the two layer in the bulk unit cell. The atom positions and lattice constants are optimized under $Pmn2_{1}$ symmetry.}
			\begin{tabular}{l r r r}
				\hline\hline
				\multicolumn{4}{c}{\textbf{A. Lattice parameters}} \\
				\hline
				\textbf{Basis vector}   & $x$ [{\AA}] & $y$ [{\AA}] & $z$ [{\AA}] \\
				$\vec{a}$ & $3.71915$ & $0.00000$ & $0.00000$ \\
				$\vec{b}$ & $0.00000$ & $12.4221$ & $0.00000$ \\
				$\vec{c}$ & $0.00000$ & $0.00000$ & $23.9814$ \\
				\hline
				\textbf{Axis angle}    & $\alpha$ [$^{\circ}$] & $\beta$ [$^{\circ}$] & $\gamma$ [$^{\circ}$] \\
				& $90$ & $90$ & $90$ \\
				\hline \hline
				\multicolumn{4}{c}{\textbf{B. Atomic Wyckoff positions}} \\
				\hline 
				Ta & $0$ & $-0.05312$ & $0.08669$ \\
				Ta & $1/2$ & $0.26344$ & $0.07862$ \\
				Rh & $0$ & $0.46000$ & $0.08401$ \\
				Rh &  $1/2$ & $-0.24965$ & $0.08126$ \\
				Te & $1/2$ & $0.06155$ & $0.02046$ \\
				Te & $0$ & $-0.19740$ & $0.00513$ \\
				Te & $0$ & $-0.35181$ & $0.13854$ \\
				Te & $1/2$ & $0.40777$ & $0.16015$ \\
				Te & $1/2$ & $-0.43785$ & $0.02673$ \\
				Te & $0$ & $0.31959$ & $-0.00074$ \\
				Te & $0$ & $0.14877$ & $0.14484$ \\
				Te & $1/2$ & $-0.10929$ & $0.16604$ \\
				\hline\hline
			\end{tabular}
			\label{tab:mono}
		\end{center}
	\end{table}
	
	\clearpage
	\newpage
	
	\begin{table}
		\setlength{\tabcolsep}{20pt}
		\caption{\small Crystal structure of bilayer {\trt}. It was constructed from the bulk unit cell by inserting a large vacuum between the bilayer in the bulk unit cell. The atom positions and lattice constants are not optimized.}
		\centering
		\begin{tabular}{l r r r}
			\hline\hline
			\multicolumn{4}{c}{\textbf{A. Lattice parameters}} \\
			\hline
			\text{Basis vector}   & $x$ [{\AA}] & $y$ [{\AA}] & $z$ [{\AA}] \\
			$\vec{a}$ & $3.75672$ & $0.00000$ & $0.00000$ \\
			$\vec{b}$ & $0.00000$ & $12.5476$ & $0.00000$ \\
			$\vec{c}$ & $0.00000$ & $0.00000$ & $30.6581$ \\
			\hline
			\text{Axis angle}    & $\alpha$ [$^{\circ}$] & $\beta$ [$^{\circ}$] & $\gamma$ [$^{\circ}$] \\
			& $90$ & $90$ & $90$ \\
			\hline \hline
			\multicolumn{4}{c}{\textbf{B. Atomic Wyckoff positions}} \\
			\hline 
			Ta & $0$ & $0.05312$ & $0.28409$\\
			Ta &  $1/2$ & $-0.05312$ & $0.06936$\\
			Ta  & $0$ & $0.27076$ & $0.06376$\\
			Ta  & $1/2$ & $ -0.27076$ & $0.27848$\\
			Rh & $0$ & $-0.46534$ & $0.28204$\\
			Rh & $1/2$ & $0.46534$ & $0.06731$\\
			Rh & $0$ & $-0.24412$ & $0.06376$\\
			Rh & $1/2$ & $0.24412$ & $0.27848$\\
			Te & $0$ & $0.06713$ & $0.01782$\\
			Te & $1/2$ & $ -0.06694$ & $0.23429$\\
			Te & $0$ & $0.19176$ & $0.21742$\\
			Te & $1/2$ & $-0.19081$ & $0.00238$\\
			Te & $0$ & $0.34590$ & $0.32485$\\
			Te & $1/2$ & $-0.34578$ & $0.109175$\\
			Te & $0$ & $0.41368$ & $0.12874$\\
			Te & $1/2$ & $-0.41280$ & $0.34376$\\
			Te & $0$ & $-0.43343$ & $0.02092$\\
			Te & $1/2$ & $0.43304$ & $0.23636$\\
			Te & $0$ & $-0.32326$ & $0.21514$\\
			Te & $1/2$ & $0.32340$ & $0$\\
			Te & $0$ & $-0.15175$ & $0.33065$\\
			Te & $1/2$ & $0.15042$ & $0.11414$\\
			Te & $0$ & $-0.10495$ & $0.13257$\\
			Te & $1/2$ & $0.10627$ & $0.34764$\\
			\hline\hline
		\end{tabular}
		\label{tab:bily}
	\end{table}
	
	\clearpage
	\newpage
	
	\begin{table}
		\setlength{\tabcolsep}{20pt}
		\caption{Crystal structure of bulk TaRhTe$_4$}
		\centering
		\begin{tabular}{l r r r}
			\hline\hline
			\multicolumn{4}{c}{\textbf{A. Lattice parameters}} \\
			\hline
			\text{Basis vector}   & $x$ [{\AA}] & $y$ [{\AA}] & $z$ [{\AA}] \\
			$\vec{a}$ & $3.75672$ & $0.00000$ & $0.00000$ \\
			$\vec{b}$ & $0.00000$ & $12.5476$ & $0.00000$ \\
			$\vec{c}$ & $0.00000$ & $0.00000$ & $13.1660$ \\
			\hline
			\text{Axis angle}    & $\alpha$ [$^{\circ}$] & $\beta$ [$^{\circ}$] & $\gamma$ [$^{\circ}$] \\
			& $90$ & $90$ & $90$ \\
			\hline \hline
			\multicolumn{4}{c}{\textbf{B. Atomic Wyckoff positions}} \\
			\hline 
			Ta & $0$ & $0.05311$ & $-0.19200$\\
			Ta & $0$ & $0.27076$ & $0.29495$\\
			Rh & $0$ & $-0.46534$ & $-0.19677$\\
			Rh & $0$ & $-0.24412$ & $ 0.29496$\\
			Te & $0$ & $0.06743$ & $0.19535$\\
			Te & $0$ & $0.19140$ & $-0.34448$\\
			Te & $0$ & $0.34526$ & $-0.10208$\\
			Te & $0$ & $0.41389$ & $0.44296$\\
			Te & $0$ & $-0.43341$ & $0.19958$\\
			Te & $0$ & $ -0.32390$ & $-0.34990$\\
			Te & $0$ & $-0.15087$ & $-0.08964$\\
			Te & $0$ & $-0.10633$ & $ 0.45250$\\
			\hline\hline
		\end{tabular}
		\label{tab:blk}
	\end{table}
	
	\clearpage
	\newpage
	
	\section{\label{sec:App_mband}Band structure of monolayer {\trt} with bulk lattice constants. }
	
	In Supplementary Fig.~\ref{figS4}(a), we show the band structure of monolayer TaRhTe$_4$ without SOC. The Dirac cone can be found along path Y-S akin to the case with optimized lattice constants, but the band along $\Gamma$-X touches Fermi level showing the difference.
	With SOC, in Fig.~\ref{figS4}(b), the band gap $E_{g}=76$ meV is larger than the optimized results ($65$ meV).
	\begin{figure}[ht!]
		\centering
		\includegraphics[width=0.99\columnwidth]{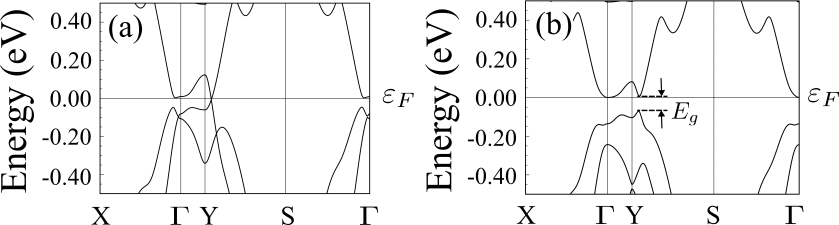}
		\caption{Band structure of monolayer TaRhTe$_4$ (a) without SOC and (b) with SOC.
		}
		\label{figS4}
	\end{figure}
	\clearpage
	\newpage
	
	\section{\label{sec:exfoliation}Exfoliation energy of monolayer and bilayer {\trt}. }
	
\begin{figure}[ht!]
		\centering
		\includegraphics[width=0.6\columnwidth]{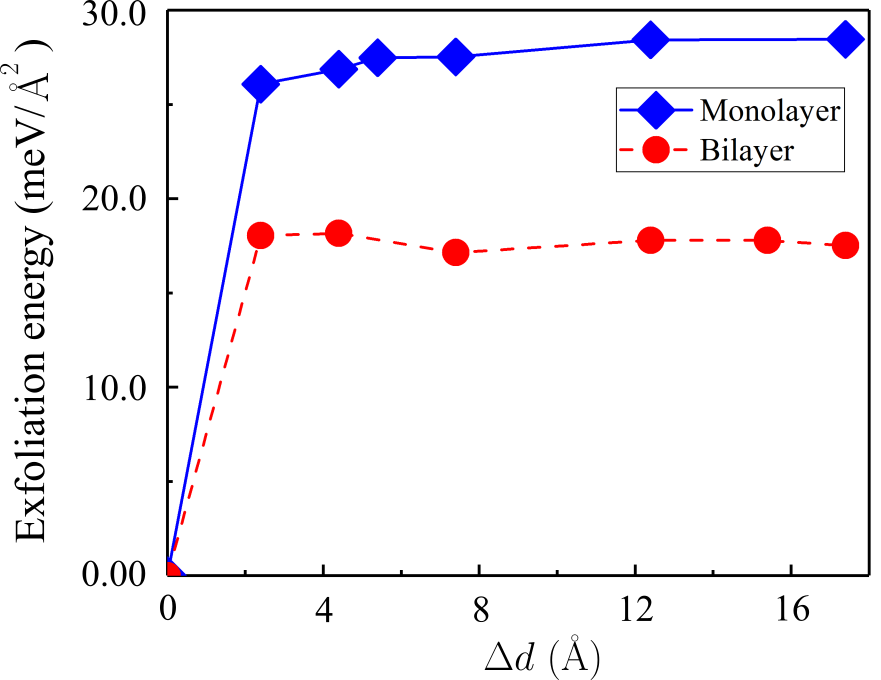}
		\caption{Exfoliation energy of monolayer and bilayer {\trt}.
		}
		\label{exf}
	\end{figure}

	Since each unit cell of the bulk {\trt} has two layers, we increase the interlayer distance between unit cells to approach the bilayer limit.
	To arrive at the monolayer limit, it is necessary to increase both the interlayer distance between unit cells and the interlayer distance within unit cells.
	The exfoliation energy is defined as the total energy after changing the distance between layers minus the total energy of the bulk in the equilibrium state.
	In Fig.~\ref{exf}, we show the exfoliation energy as a function of $\Delta d$, where $\Delta d=d-d_0$ denotes the difference of interlayer distance and equilibrium distance. 
	
	The exfoliation energy for monolayer {\trt} is $28.5$ meV/{\AA}$^2$, while the energy for bilayer {\trt} is $17.5$
	meV/{\AA}$^2$. In addition, we exfoliate graphene from graphite in the same way for reference. We obtain the exfoliation energy of graphene to be $13$ meV/{\AA}$^2$, which is slightly smaller than the result ($21$ meV/{\AA}$^2$) in another calculation.\cite{exfogra}
	
	\clearpage
	\newpage
	
	\section{\label{sec:App_phono}Phonon spectrum of monolayer \texorpdfstring{TaRhTe4}{\trt}.}
\begin{figure}[ht!]
		\centering
		\includegraphics[width=0.6\columnwidth]{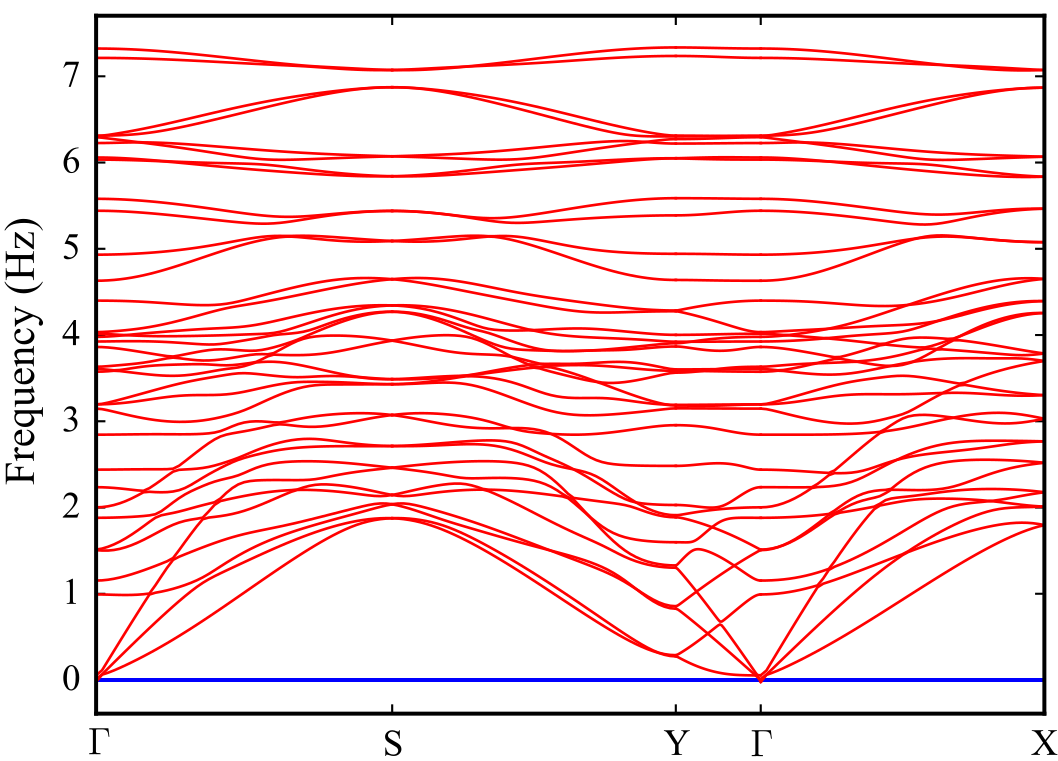}
		\caption{Phonon dispersion spectrum of monolayer TaRhTe$_4$.
		}
		\label{figS5}
	\end{figure}

	Figure~\ref{figS5} serves as confirmation of the dynamic stability of monolayer {\trt}. It displays the phonon spectrum of monolayer structure. The acoustic branches and the other optical branches are clearly separated from each other in the vicinity of the $\Gamma$ point.\cite{D2QI01608G} There are phonon branches observed below zero frequency through the BZ.
	
	\clearpage
	\newpage
	
	\section{\label{sec:bilayer_inversion}Layer-resolved orbital contributions to the bandstructure in bilayer {\trt}. }
	
\begin{figure}[ht!]
		\centering
		\includegraphics[width=0.99\columnwidth]{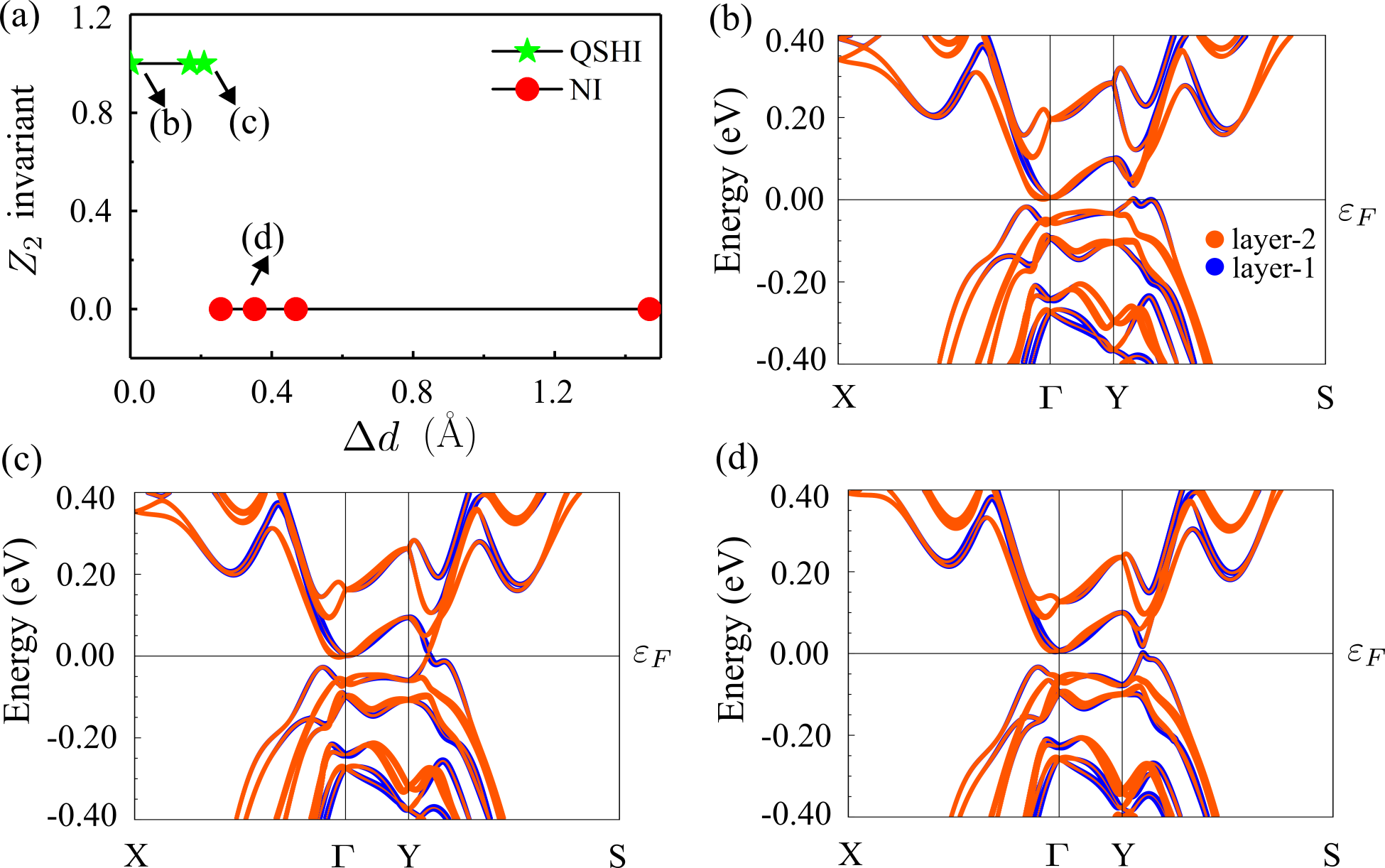}
		\caption{Layer resolved band weight of bilayer {\trt} with different layer displacements.
		}
		\label{biband}
	\end{figure}

	In order to understand the topological phase transtion in the bilayer 
	{\trt}, we present the band weight of bilayer system projected to 
	orbitals from different layers. Figure~\ref{biband}(b) is the band 
	weight of equilibrium bilayer structure. We find that the layer-1 
	dominates valence band edge (near the Y point) whereas conduction band 
	edge are governed by layer-2. Approaching the TPT point, the band gap is 
	closing near the Y point as shown in Fig.~\ref{biband}(c). After 
	crossing the phase boundary, in Fig.~\ref{biband}(d), the contribution of both the van der Waals 
	separated layers tend to equate in the conduction and valence band edges.
	\clearpage
	\newpage
	
	\section{\label{sec:App1}Edge states of monolayer and bilayer {TaRhTe$_4$}. }
	
	In Supplementary Fig.~\ref{fig8}, we show the spectra of edge states of monolayer TaRhTe$_4$ with different terminations. In Supplementary Fig.~\ref{figS3}, we show the spectra of edge states of bilayer TaRhTe$_4$.
	
	\begin{figure}[ht!]
		\centering
		\includegraphics[width=0.95\columnwidth]{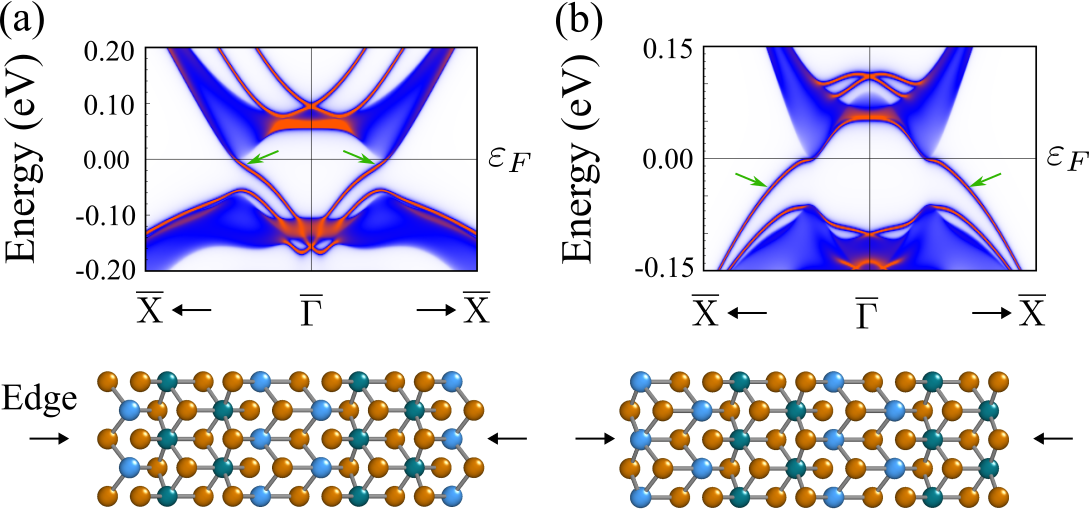}
		\caption{Edge state spectra of monolayer TaRhTe$_4$ with different terminations.
		}
		\label{fig8}
	\end{figure}
	
	\begin{figure}[ht!]
		\centering
		\includegraphics[width=0.5\columnwidth]{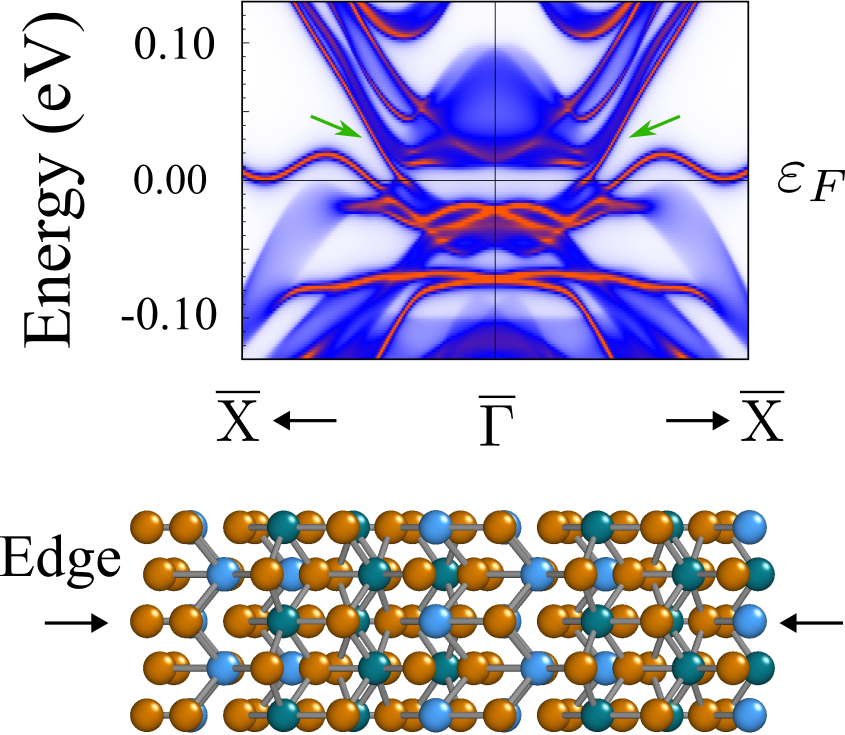}
		\caption{Edge state spectra of bilayer TaRhTe$_4$ and corresponding terminations.
		}
		\label{figS3}
	\end{figure}

	\clearpage
	\newpage
	
	\section{\label{sec:AAstack_bands}Bandstructure and Wilson loop of AA-stacked bilayer {\trt}. }
	In Supplementary Fig.~\ref{fig:bandAA}, we show the bandstructure and Wilson loop of AA-stacked TaRhTe$_4$. The interlayer distance (vertical distance of Rh atoms from different layers) is $6.358$ {\AA}.
	\begin{figure}[ht!]
		\centering
		\includegraphics[scale=0.520]{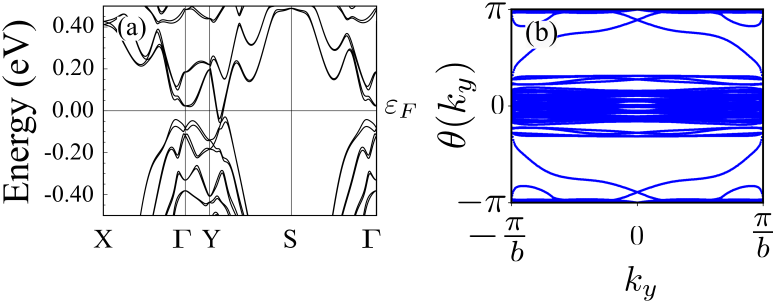}
		\caption{\small (a) Band structure of AA-stacked bilayer TaRhTe$_4$ along high symmetry points with SOC. (b) Wislon loop for AA-stacked bilayer TaRhTe$_4$.
		}
		\label{fig:bandAA}
	\end{figure}
	
	\clearpage
	\newpage
	\section{\label{sec:App_bulk_bands}Bandstructure of bulk 3D {\trt}. }
\begin{figure}[ht!]
		\centering
		\includegraphics[scale=0.520]{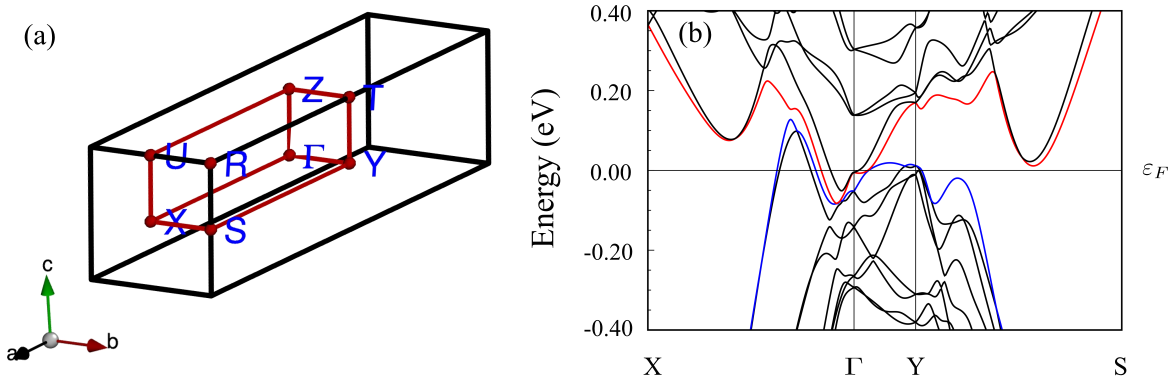}
		\caption{\small (a) Brillouin zone of bulk \trt along with the high symmetry points. (b) Band structure of 3D TaRhTe$_4$ along high symmetry points with SOC. The blue curve represents the highest valence band, while the red curve is the lowest conduction band.
		}
		\label{fig:bands_3d_trt}
	\end{figure}

	Supplementary Figure~\ref{fig:bands_3d_trt} shows the BZ and DFT bandstructure for bulk (3D) {\trt}. The top of the valence band (band $N$) and the bottom of the conduction band (band $N+1$), used for obtaining the WPs, are highlighed in blue and red color, respectively.
	
\end{document}